\documentclass[aps,pre,twocolumn,groupedaddress,longbibliography]{revtex4-1}
\usepackage{graphicx,amsmath,amssymb,color}
\begin{document}

\title{Ultraslow settling kinetics of frictional cohesive powders}
\author{Kai Nan}
\author{Robert S. Hoy}
\email{rshoy@usf.edu}
\affiliation{Department of Physics, University of South Florida, Tampa, FL 33620}
\date{\today}
\begin{abstract}
Using discrete element method simulations, we show that the settling of  frictional cohesive grains under ramped-pressure compression exhibits strong history dependence and slow dynamics that are not present for grains that lack either cohesion or friction.
Systems prepared by beginning with a dilute state and then ramping the pressure to a small positive value $P_{\rm final}$ over a time $\tau_{\rm ramp}$ settle at packing fractions given by an inverse-logarithmic rate law, $\phi_{\rm settled}(\tau_{\rm ramp}) = \phi_{\rm settled}(\infty) +  A/[1 + B\ln(1 + \tau_{\rm ramp}/\tau_{\rm slow})]$.
This law is analogous to the one obtained from classical tapping experiments on noncohesive grains, but crucially different in that $\tau_{\rm slow}$ is set by the slow dynamics of structural void stabilization rather than the faster dynamics of bulk densification.
We formulate a kinetic free-void-volume theory that predicts this  $\phi_{\rm settled}(\tau_{\rm ramp})$, with $\phi_{\rm settled}(\infty) = \phi_{\rm ALP}$ and $A = \phi_{\rm settled}(0) - \phi_{\rm ALP}$, where $\phi_{\rm ALP} \equiv .135$ is the ``adhesive loose packing'' fraction found by Liu \textit{et al.} [W.\ Liu, Y.\ Jin, S. Chen, H.\ A.\ Makse and S.\ Li, \textit{Soft Matt.} \textbf{13}, 421 (2017)].
\end{abstract}
\maketitle

The structure of granular solids is famously preparation-protocol-dependent.
For example, mechanical excitation by periodic tapping makes samples' packing fractions increase logarithmically in time:  
\begin{equation}
\phi(t) = \phi(\infty) - \displaystyle\frac{A}{1 + B\ln(1 + t/\tau_{\rm slow})},
\label{eq:phitap}
\end{equation}
where $A$, $B$, and $\tau_{\rm slow}$ depend on the sample-preparation and tapping protocols in addition to the intergrain interactions \cite{knight95,richard05}.
This density increase is directly analogous to, but typically far greater in extent than, the density increase experienced by aging thermal glasses \cite{kovacs63};
both arise from the slow, activated dynamics of systems traversing the rugged energy landscapes that are a common feature of thermal glasses and granular materials \cite{debenedetti01,liu10}.
Cohesive interactions greatly slow the dynamics of viscous liquids \cite{berthier09c}, and frictional interactions greatly slow the dynamics of granular solids \cite{campbell90,tencate00}.
One might expect that the combination of frictional and cohesive interactions will produce a further dynamical slowdown, and indeed it does. 
In particular, the combination of cohesive interactions, rolling, sliding, and twisting friction can arrest compaction entirely -- at least on human time scales -- by mechanically stabilizing large ``structural'' voids within marginally jammed packings \cite{kadau03,bartels05,vandewalle07,gilabert07,gilabert08,kadau10}.

As a consequence, unlike their frictionless or purely-repulsive counterparts, frictional cohesive granular solids can be prepared with a very wide range of densities.
For example, the Hausner ratio $H = \rho_{\rm tapped}/\rho_{\rm settled}$ \cite{hausner67}, where $\rho_{\rm settled}$ is the density obtained by pouring grains into a container and $\rho_{\rm tapped}$ is the density obtained in the long-time limit of a tapping experiment, is a commonly employed measure of powder flowability.
$H$ is also a measure of the range of jamming densities $\phi_J$ obtainable via different preparation protocols, i.e.\ different protocols will produce $\phi_{\rm min} \leq \phi_J \leq \phi_{\rm max} \equiv H \phi_{\rm min}$.
$H$ has long been known to increase with decreasing particle size, approaching 4 for micron-size grains, because smaller grains are more cohesive than their larger counterparts \cite{carson98,blum04}.  
More recently it has been explicitly shown that $H$ values for fixed-size grains increase rapidly with both cohesion and friction \cite{vandewalle07,fiscina10}, and recent simulations that established an equation of state for random sphere packings \cite{liu15,liu17} suggest that spherical grains' $H \to H_{\rm max} \simeq 3.8$ in the limit of strong cohesion and friction.

Using $H$ as a measure of powder flowability is often criticized on the grounds that both $\rho_{\rm tapped}$ and $\rho_{\rm settled}$ are preparation-protocol dependent \cite{santomaso03}; in general, reproducible values of $H$ are obtained only when highly specific standardized procedures are followed \cite{astmb212}.
The interplay of cohesion and friction in determining the history-dependence of both ``static'' macroscopic quantities like $\rho_{\rm settled}$ and microscopic (grain-level) structure in these powders remains poorly understood and the subject of active study \cite{traina13,schmidt20,lemaitre21}.
In particular, while the logarithmically slow densification of noncohesive and frictionless cohesive granular materials has been semiquantitatively explained by kinetic free-volume theories  \cite{boutreux97,bennaim98,edwards98,hao15}, microscopic-physics-based theories that accurately predict the preparation-protocol-dependent $\phi(t)$  [including $\phi(0) \equiv \phi_{\rm settled}$] for frictional cohesive powders have yet to be developed, and doing so is very challenging owing to additional complications associated with the abovementioned mechanically stable structural voids.
Developing such theories could prove useful for applications ranging from avalanche prevention \cite{mede20} to  pharmaceuticals \cite{goh18} to additive manufacturing \cite{vock19}.

In this Letter, we use discrete element method (DEM) simulations to examine how the structure of marginally jammed systems of grains with varying degrees of friction and cohesion depends on the compression protocol used to prepare them. 
We compare results for model systems with four types of intergrain interactions: (1) no friction or cohesion, (2) all three types of friction (sliding, rolling, and twisting) but no cohesion, (3) cohesion but no friction, and (4) both cohesion and friction.  
The settled packing fractions of systems prepared by beginning with a dilute state and then linearly ramping the pressure to a fixed, small value $P_{\rm targ}$ over a time $\tau_{\rm ramp}$ decrease as cohesion and friction are increased, ranging from the canonical random-close-packed value ($\phi_{\rm RCP} = 0.646$  \cite{zhang05,ciamarra10b}) for model 1 to as low as $0.35$ for model 4.  
While these $\phi_{\rm settled}$ are almost independent of $\tau_{\rm ramp}$ for models 1-3, they decrease substantially with increasing $\tau_{\rm ramp}$ for model 4, reaching their asymptotic low-rate limit at a $\tau_{\rm ramp}$ that is many orders of magnitude larger than the corresponding values for models 1-3.

This behavior is the opposite of the usual glass-jamming paradigm \cite{debenedetti01,chaudhuri10}, in which thermal glasses and granular materials end up with higher densities when they are more slowly cooled or compressed.
We find that the rate-dependence of model 4's $\phi_{\rm settled}$ is described by 
\begin{equation}
\phi_{\rm settled}(\tau_{\rm ramp}) = \phi_{\rm settled}(\infty) + \displaystyle\frac{A}{1 + B\ln{\left(1 + \tau_{\rm ramp}/\tau_{\rm slow}\right)}},
\label{eq:pjoftr}
\end{equation}
and argue that the difference leading to the crucial change in sign (from $-$ to $+$) is that while the $\tau_{\rm slow}$ in Eq.\ \ref{eq:phitap} is set by the slow dynamics of densification  \cite{knight95,richard05}, the $\tau_{\rm slow}$ in Eq.\ \ref{eq:pjoftr} is set by the even slower dynamics of \textit{structural void stabilization}.
Then we formulate a kinetic free-void-volume theory (similar in spirit to but different in several crucial details from those of Refs.\ \cite{boutreux97,bennaim98,edwards98,hao15}) that predicts this behavior, with $\phi_{\rm settled}(\infty) =  \phi_{\rm ALP}$ and  $A = \phi_{\rm settled}(0) - \phi_{\rm ALP}$, where $\phi_{\rm ALP} \equiv .135$ is the ``adhesive loose packing'' fraction found by Liu \textit{et al.}\ \cite{liu15,liu17}.

Our simulations aim to implement realistic viscoelastic, cohesive and frictional interactions in a way that is computationally cheap enough to allow us to simulate large systems over long timescales. 
Therefore we choose to employ the Hertzian variant of the widely-used Rognon potential \cite{rognon06,rognon08} for the conservative pair interactions.
A standard radial damping force  \cite{brilliantov96} is added to capture viscous dissipation.
Sliding friction is implemented using the widely used linear-history model \cite{silbert01}, while rolling and twisting friction are implemented using the same methods as Santos \textit{et al.}\ \cite{santos20,luding08}.
Ref.\ \cite{santos20} showed that this combination of intergrain interactions accurately reproduces the packing fractions and coordination numbers found in typical experiments \cite{scott69,jerkins08,farrell10,schmidt20}.
Since we wish to consider the limit of strong friction in this study, we set the sliding, rolling, and twisting friction coefficients to $0.5$.
All interactions are described in detail in the Supplementary Materials; all quantities discussed below are expressed in dimensionless units.

\begin{figure}[htbp]
\includegraphics[width=2.775in]{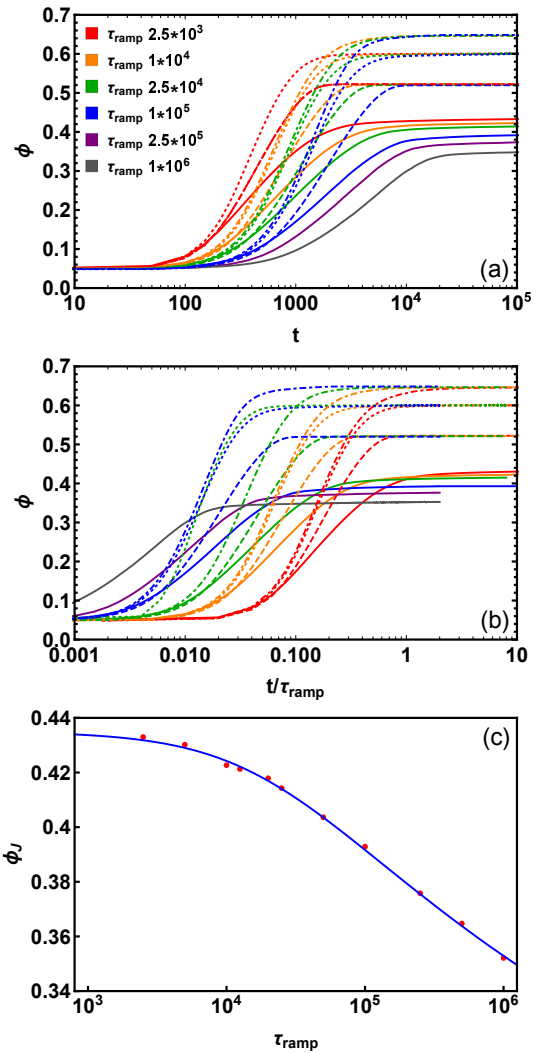}
\caption{Influence of intergrain interactions and preparation protocol on powder settling.  
Panels (a-b) respectively show $\phi(t)$ and  $\phi(t/\tau_{\rm ramp})$ for a wide range of $\tau_{\rm ramp}$.  Dot-dashed, dotted, dashed and solid curves respectively show results for models 1-4.
All results for models 1-3 are consistent with many previous studies, e.g.\ Refs.\ \cite{ohern03,silbert10,liu15,liu17,santos20}.
Panel (c) shows the settled densities $\phi_{\rm settled} = \phi(10 \tau_{\rm ramp})$ for frictional cohesive grains (model 4).  Red symbols show simulation data and the blue curve shows Eq.\  \ref{eq:pjoftr} with $\phi_{\rm settled}(\infty) = 0.135$, $A = 0.300$, $B = 0.098$, and $\tau_{\rm slow} = 2.2\cdot 10^4$. }
\label{fig:phivst100K}
\end{figure}

DEM simulations are performed using LAMMPS \cite{thompson22}.
Following Ref.\ \cite{santos20}, we begin by placing $N = 10^5$ particles randomly within a periodic cubic simulation cell of volume  $V = 39N\pi/(125\phi_{\rm init})$, where $\phi_{\rm init} = .05$ is the initial packing fraction.
Then a slow pushoff is run (at constant $\phi$) to eliminate high-energy particle overlaps; the liberated energy is removed by damping the particle velocities until an athermal state is obtained.
After the pushoff is completed, ``settled'' states are prepared using a procedure that mimics pouring a powder into a container at a rate $R_P \sim \tau_{\rm ramp}^{-1}$, but removes complications associated with pouring experiments' anisotropic ``external'' forces (i.e.\ gravity and the container walls).
We ramp the applied hydrostatic pressure from $0$ to $P_{\rm targ} = 10^{-3}$ over a time $\tau_{\rm ramp}$ and afterwards hold it constant for least another $10^5$ time units.
In other words, the applied stresses along the $x$, $y$ and $z$ directions are $\sigma_{xx} = \sigma_{yy} = \sigma_{zz}  = P_{\rm targ} t/\tau_{\rm ramp}$ for $t \leq \tau_{\rm ramp}$, and $\sigma_{xx} = \sigma_{yy} = \sigma_{zz}  = P_{\rm targ}$ for $t \geq \tau_{\rm ramp}$ \cite{footphifluct,foottriclinic}.
Since $P_{\rm targ} = 10^{-3}$ is large enough for the employed Nose-Hoover barostat to be effective yet small enough to minimize plastic consolidation  \cite{gilabert07,gilabert08}, we define all systems' $\phi_{\rm settled}$ as their $\phi(\tau_{\rm ramp} + 10^5)$.
This definition closely corresponds to the $\phi_{\rm settled}$ that could be measured after the termination of a pouring experiment.

Figure \ref{fig:phivst100K} shows the $\tau_{\rm ramp}$-dependent responses for all four models.
As expected, results for repulsive frictionless spheres (model 1) show negligible preparation-protocol dependence.
All systems settle at $\phi \simeq 0.646$; this density is consistent with random close packing \cite{zhang05,ciamarra10b}.
The $\phi(t)$ curves nearly collapse when replotted vs. $t/\tau_{\rm ramp}$, at least for $t/\tau_{\rm ramp} > 1$ [panel (b)].
For smaller  $t/\tau_{\rm ramp}$, $\phi$ increases with increasing $\tau_{\rm ramp}$ owing to well-understood kinetic effects associated with the hard sphere glass transition \cite{speedy98}.
Comparable preparation-protocol independence of the final jammed states and collapse of the $\phi(t)$ curves occurs for systems with friction but no cohesion (model 2) or cohesion but no friction (model 3), but at lower $\phi_{\rm settled}$.
Model 2 systems have $\phi_{\rm settled} \simeq 0.60$, which is consistent with the results of Santos \textit{et al.}\ \cite{santos20} for our employed value of $P_{\rm targ}$. 
Model-3 systems have $\phi_{\rm settled} \simeq 0.52$, which is consistent with adhesive close packing \cite{liu15,liu17} in the presence of the finite-range attractive intergrain interactions (which favor finite particle overlap) employed in this study.
These models do not show any evidence of compaction dynamics that are significantly slower than those of model 1.
Indeed their $\phi(t)$ actually converge slightly faster, perhaps because their $\phi_{\rm settled}$ are lower and hence their nearly-settled states have more free volume.

Results for systems with both cohesion and friction (model 4) are radically different.
Their $\phi(t/\tau_{\rm ramp})$ increase (decrease) monotonically with increasing $\tau_{\rm ramp}$ for $t/\tau_{\rm ramp} \ll 1$ ($t/\tau_{\rm ramp} \gtrsim 1$), and are still increasing logarithmically slowly at $t = 10\tau_{\rm ramp}$  in a manner reminiscent of tapping experiments \cite{knight95,richard05}, 
but show no evidence of convergence towards history-independent values.
As shown in panel (c), our results can be well fit by Eq.\ \ref{eq:pjoftr}.
We assumed $\phi_{\rm settled}(\infty) = \phi_{\rm ALP}$ since this is the packing fraction expected in the limit of large system size and slow compression for systems with very strong cohesion and friction \cite{liu17}.
$A \simeq 0.300$ is a fitting parameter capturing the range of $\phi$ obtainable as pressure ramping varies from infinitely fast to infinitely slow.
$B \simeq 0.098$ is a fitting parameter capturing the relative importance of the logarithmic term \cite{hao15}.
Finally, $\tau_{\rm slow} \simeq 2.2\cdot 10^4$ is a time scale capturing model 4's inherently slow dynamics.

Clearly Eq.\ \ref{eq:pjoftr} is directly analogous to Eq.\ \ref{eq:phitap}, but with a crucial difference.
Longer tapping duration produces higher densities, whereas slower pressure ramping produces \textit{lower} densities.
The latter behavior is the opposite of the usual glass-jamming paradigm \cite{debenedetti01,chaudhuri10}, in which thermal glasses and granular materials end up with higher densities when they are more slowly cooled or compressed.
The $-$ sign between the two terms in Eq.\ \ref{eq:phitap} is associated with the slow dynamics of densification in tapped systems \cite{knight95}; comparable dynamics control densification of aging thermal glasses \cite{kovacs63}.
In contrast, as we will show below, the $+$ sign between the two terms in Eq.\ \ref{eq:pjoftr} is associated with a slow dynamics of \textit{void stabilization}.

\begin{figure}[h!]
\includegraphics[width=2.775in]{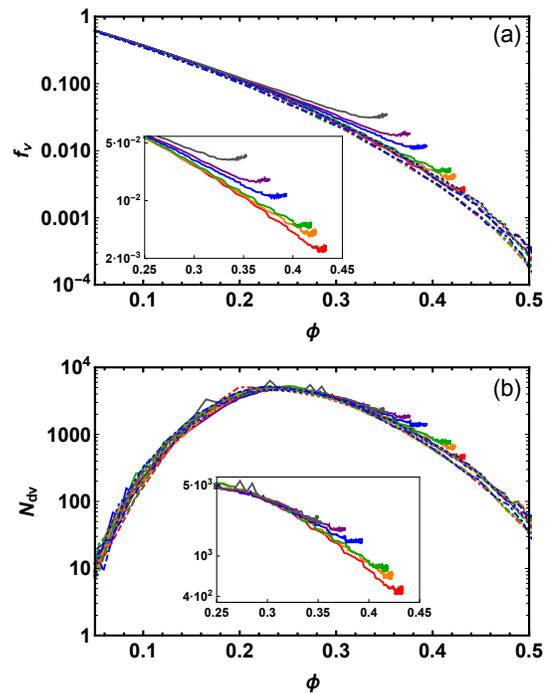}
\caption{Void volume fraction $f_{\rm v}$ [panel (a)] and number of topologically distinct voids $N_{\rm dv}$ [panel (b)].   Insets highlight the void stabilization that occurs for model 4.  All colors and line types are the same in Fig. \ref{fig:phivst100K}.}
\label{fig:voidstats}
\end{figure}

We monitored void growth and coalescence by dividing the DE simulation cells into $N_{\rm c}(t) = n_x(t) \times n_y (t) \times n_z (t)$ subcells of side lengths  $\ell_x(t) = L_x(t)/\rm{floor}[L_x(t)]$,   $\ell_y(t) = L_y(t)/\rm{floor}[L_y(t)]$, and  $\ell_z(t) = L_z(t)/\rm{floor}[L_z(t)]$; here $\rm{floor}[\xi]$ rounds $\xi$ downward to the nearest integer.   
A subcell is classified as a void if it intersects no (i.e., contains no portion of any) particle cores, where the core of particle $j$ is the sphere of radius $R_j$ centered at $\vec{r}_j$.
Thus void subcells are subcells inside which at least one small particle can be placed without contacting any other particles.
The void volume fraction is defined as $f_{\rm v}(t) = N_{\rm vc}(t)/N_{\rm c}(t)$, where $N_{\rm vc}(t)$ is the total number of void subcells.
We divide these $N_{\rm vc}(t)$ void subcells into $N_{\rm dv}(t)$ distinct (topologically disconnected) voids using connected-components analysis \cite{hopcroft73}, and define structural voids as distinct voids of volume $\geq 10$.

Results for all systems are shown in Figure \ref{fig:voidstats}.
For models 1-3, $f_{\rm v}$ decreases approximately exponentially with $\phi$ and drops to zero (to within our statistical accuracy) by $\phi \simeq 0.55$.
Cohesive systems have larger $f_{\rm v}$ than their noncohesive counterparts for all $\phi$, largely because their constituent grains are more likely to form compact clusters at lower $\phi$ \cite{lois08,koeze18,koeze20}, but the slopes $d[\ln(f_{\rm v})]/d\phi$ are similar for all three models.
As compression continues, $-d[\ln(f_{\rm v})]/d\phi$ increases as void-filling becomes more coherent, i.e.\ as free volume decreases and particles are increasingly likely to get pushed into empty regions by their interactions with other particles.
Results for $N_{\rm dv}(\phi)$ show complementary trends.  
As compression proceeds, $N_{\rm dv}$ initially increases as large voids are split into smaller ones (recall that a homogeneous system in the low-$\phi$ limit would have $N_{\rm dv} = 1$), then decreases as these small voids get filled.
For $\phi \gtrsim 0.4$, most voids consist of only one or two subcells, so $N_{\rm dv}$ roughly tracks $f_{\rm v}$.

Model 4 systems' void statistics follow similar trends at low $\phi$.
Their $f_{\rm v}(\phi)$ are slightly higher than their model-3 counterparts, presumably because the compact clusters they form are mechanically
stabilized by their frictional interactions and hence are more likely to grow as compression continues \cite{dong22}.
As compression continues, however, the behavior of these systems again becomes qualitatively different from that of models 1-3.
Both $f_{\rm v}(\phi)$ and $N_{\rm dv}(\phi)$ begin rising substantially above the common exponential trends, at packing fractions $\phi_{\rm vso}$ that decrease rapidly with increasing $\tau_{\rm ramp}$.
Evidently these $\phi_{\rm vso}(\tau_{\rm ramp})$ correspond to the onset of structural voids' mechanical stabilization, with lower $\phi_{\rm vso}$ leading to larger final  $f_{\rm v}$ and $N_{\rm dv}$ and therefore also to lower $\phi_{\rm settled}$.

\begin{figure}[htbp]
\includegraphics[width=3in]{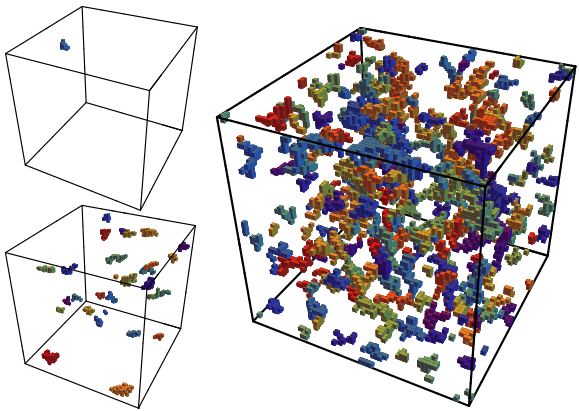}
\caption{Structural voids in model 4's final settled states.  
The upper left, lower left, and right images respectively show results for $\tau_{\rm ramp} = 10^4$, $10^5$ and $10^6$; different colors indicate topologically distinct voids.}
\label{fig:voidpic}
\end{figure}

Visualizing these voids both illustrates the above arguments and reveals a feature that was not apparent from the $f_{\rm v}$ and $N_{\rm dv}$ data alone.
Figure \ref{fig:voidpic} shows how increasing $\tau_{\rm ramp}$ \textit{qualitatively} alters the final structural-void geometry. 
For $\tau_{\rm ramp} = 10^4$, only one small structural void (of volume 11) is present in the settled configuration.
In contrast, the settled configurations for $\tau_{\rm ramp} = 10^5$ ($\tau_{\rm ramp} = 10^6$) contain 27 (218) structural voids, with volumes as large as $48$ (261).
Thus larger $\tau_{\rm ramp}$ lead not only to larger $f_{\rm v}$ and correspondingly lower $\phi_{\rm settled}$, but also to dramatic increases in the number and maximum size of structural voids, and consequently in the settled states' spatial heterogeneity.
Note that the final settled states for models 1-3 have \textit{no} structural voids for the range of $\tau_{\rm ramp}$ considered here, and their large-scale spatial heterogeneity
[as indicated, e.g., by the low-$q$ limit of the static structure factor $S(q)$] is $\tau_{\rm ramp}$-independent.

The first  theories that successfully explained the logarithmically slow increase of $\phi(t)$ during tapping experiments (Eq.\ \ref{eq:phitap}) did so by noting that free volume decreases exponentially, and therefore the characteristic time between relaxation events that lead to further densification increases exponentially, with increasing $\phi$  \cite{boutreux97,bennaim98,edwards98}. 
In the same spirit, we postulate that the kinetic effects of increasing the void volume fraction $\phi_v \equiv 1-\phi$ towards $1 - \phi_{\rm ALP}$  in a settling frictional cohesive powder are comparable to the effects of increasing $\phi$ towards $\phi_{\rm RCP}$ in a tapped frictionless noncohesive powder.
In other words, we assume that the ``free void volume'' vanishes for $\phi < \phi_{\rm ALP}$ because $\phi$ cannot be reduced any further without destabilizing the powder, and therefore the characteristic time for assembly processes that will produce settled packings with $\phi = \phi_{\rm ALP}$ is astronomical, but that this time \textit{decreases} exponentially with increasing $\phi$.

Assuming that $\tau_{\rm slow}^{-1}$ is the ``attempt rate'' for processes that form a mechanically stable settled sample, replacing the tapping-experiment duration $t$ (Eq.\ \ref{eq:phitap}) with the pouring-experiment duration $\tau_{\rm ramp}$, and adapting the procedure used in Section 2.1 of Ref.\ \cite{hao15} to the abovementioned assumptions about free void volume leads to the prediction
\begin{equation}
\exp\left[ \displaystyle\frac{\phi_{\rm settled}(0) - \phi_{\rm ALP}}{\phi_{\rm settled}(\tau_{\rm ramp}) - \phi_{\rm ALP}} \right]  = \exp(1) \left( 1 + \displaystyle\frac{\tau_{\rm ramp}}{\tau_{\rm slow}} \right)^B,
\label{eq:exptau}
\end{equation}
where $\phi_{\rm settled}(0)$ is the packing fraction obtained in the fast-pouring limit where minimal aggregation and compact-cluster stabilization occurs prior to settling \cite{dong22}, and $B$ is a free parameter that depends on factors such as the grains' size distribution and stiffness.
Rearranging Eq.\ \ref{eq:exptau} leads to the rate law
\begin{equation}
\phi_{\rm settled}(\tau_{\rm ramp}) = \phi_{\rm ALP} + \displaystyle\frac{\phi_{\rm settled}(0) - \phi_{\rm ALP}}{1 + B\ln\left( 1+  \tau_{\rm ramp}/\tau_{\rm slow} \right)}.
\label{eq:newfvkinetics}
\end{equation}
As illustrated in Fig.\ \ref{fig:phivst100K}(c), Equation \ref{eq:newfvkinetics} accurately describes model 4's $\phi_{\rm settled}(\tau_{\rm ramp})$. 
Notably, it predicts that frictional cohesive powders have \textit{ultraslow} settling kinetics in the sense that their $\phi_{\rm settled}$ continues decreasing steadily with increasing $\tau_{\rm ramp}$ even with $\tau_{\rm ramp}$ is \textit{very} large \cite{slowparking}.
Comparably slow kinetics are predicted by some ``parking lot'' models of granular compaction -- see e.g.\ Fig.\ 2 of Ref.\ \cite{silbani16} -- but these models have not yet been adapted to capture the consequences of structural void stabilization.

Equation \ref{eq:newfvkinetics} should also predict the settling kinetics of real powders in the limit of strong intergrain cohesion and friction, e.g.\ Geldart Group C \cite{geldart73} powders with average grain size $\lesssim 10\mu m$, when the pouring height is small or the settling takes place in a gas-fluidized bed.
A direct experimental test of its validity could potentially be performed by starting with a well-fluidized deagglomerated micropowder \cite{blum04,raganati18}, and then comparing the $\phi_{\rm settled}$ obtained after imposing a variety of gas-flow histories $v_g(t) = v_g(0)[1 - t/\tau_{\rm ramp}]$, where $v_g(0)$ is above  the critical fluidization velocity $v_c$ \cite{valverde01} and the set of $\tau_{\rm ramp}$ employed spans at least $\sim\! 3$ orders of magnitude.
Performing such experiments and better understanding the ultraslow kinetics of frictional cohesive powder settling could ultimately help develop more-robust processing strategies for micropowders; developing such strategies is a major current challenge in the pharmaceutical and additive-manufacturing industries  \cite{goh18,vock19}.

We thank Andrew Santos, Ishan Srivastava, Abhinendra Singh, and Corey O'Hern for helpful discussions.
This material is based upon work supported by the National Science Foundation under Grant DMR-2026271.


\begin{thebibliography}{58}
\bibitem [{foo({\natexlab{b}})}]{foottriclinic}%
  \BibitemOpen
  \href@noop  \bibinfo {Throughout these
  simulations, the periodic-cell side lengths $L_x(t)$, $L_y(t)$ and $L_z(t)$
  vary independently, so the final simulation cell is (in general)
  orthorhombic. Ref.\ \cite{santos20} showed that the stability of marginally
  jammed states against shear is optimized by compression protocols that also
  maintain the three shear stresses $\sigma_{xy}$, $\sigma_{xz}$, and
  $\sigma_{yz}$ at zero during compression and thus end up with triciinic
  simulation cells. We performed compression runs for all four models using
  such a protocol, and found that in all cases differences in the results
  discussed below were negligible.}\BibitemShut {Stop}%
\bibitem [{foo({\natexlab{b}})}]{slowparking}%
  \BibitemOpen
  \href@noop  \bibinfo {For example, it predicts that model 4's $\phi_{\rm settled}$ decreases from $\sim\! 0.282$ to $\sim\!  0.245$ ($\sim\! 0.223$ to $\sim\!  0.209$) over the range  $10^9 \leq \tau_{\rm ramp} \leq 10^{12}$ ($10^{15} \leq \tau_{\rm ramp} \leq 10^{18}$).}\BibitemShut {Stop}%
\end{thebibliography}

%

\end{document}